\documentclass[conference]{IEEEtran}
\IEEEoverridecommandlockouts
\def\DRAFT
\usepackage{cite}
\usepackage{amsmath,amssymb,amsfonts}
\usepackage{algorithmic}
\usepackage{graphicx}
\usepackage{textcomp}
\usepackage{xcolor}
\usepackage{hyperref}
\usepackage{multirow}
\usepackage{nicefrac}
\usepackage{balance}
\usepackage{hyperref}
\def\BibTeX{{\rm B\kern-.05em{\sc i\kern-.025em b}\kern-.08em
    T\kern-.1667em\lower.7ex\hbox{E}\kern-.125emX}}
\usepackage{bm}
\usepackage{mathtools}
\usepackage{amsmath}
\usepackage{multirow}
\usepackage{tabularx}
\usepackage{balance}
\usepackage{enumitem}
\usepackage{booktabs}
\usepackage[english]{babel}
\usepackage[autostyle,english=american]{csquotes}
\usepackage{graphics} %
\usepackage{threeparttable}
\usepackage{tikz}
\usetikzlibrary{automata,arrows,calc,positioning}

\ifx\DRAFT\fi

\begin{document}

\title{Robust Heart Rate Detection via \\ Multi-Site Photoplethysmography
{}
}

\author{
\IEEEauthorblockN{Manuel Meier and Christian Holz}
\IEEEauthorblockA{Department of Computer Science\\
ETH Zürich, Switzerland%
}
}

\maketitle

\begin{abstract}
Smartwatches have become popular for monitoring physiological parameters outside clinical settings.
Using reflective photoplethysmography (PPG) sensors, such watches can non-invasively estimate heart rate (HR) in everyday environments and throughout a patient's day. 
However, achieving consistently high accuracy remains challenging, particularly during moments of increased motion or due to varying device placement. 
In this paper, we introduce a novel sensor fusion method for estimating HR  that flexibly combines samples from multiple PPG sensors placed across the patient's body, including wrist, ankle, head, and sternum (chest).
Our method first estimates signal quality across all inputs to dynamically integrate them into a joint and robust PPG signal for HR estimation.
We evaluate our method on a novel dataset of PPG and ECG recordings from 14 participants who engaged in real-world activities outside the laboratory over the course of a whole day.
Our method achieves a mean HR error of 2.4\,bpm, which is 46\% lower than the mean error of the best-performing single device (4.4\,bpm, head).

\noindent
\href{https://github.com/eth-siplab/MultiSite-PPG-Robust_HeartRate}{\color{magenta}{github.com/eth-siplab/MultiSite-PPG-Robust\_HeartRate}}.
\end{abstract}


\vspace{-5mm}
\section{Introduction}

Wearable devices, smartwatches in particular, have gained popularity for monitoring physiological parameters, most commonly heart rate (HR).
For this, devices typically adopt reflective photoplethysmography (PPG) to optically and thus non-invasively sense blood volume pulses for HR estimation.

However, achieving high accuracy in optical HR estimations remains challenging~\cite{bayoumy_smart_2021}, especially in settings that lead to motion artifacts~\cite{longmore_comparison_2019}.
Amongst athletes, who subject such sensors to considerable motion, the use of ECG-based HR monitoring is prevalent, where the electrophysiological measurements with a chest strap avoid high measurement errors.
Since chest straps are more obtrusive than individual PPG sensors, researchers have sought ways to compensate for motion artifacts in PPG.
Approaches range from integrating accelerometers to detect motion~\cite{temko_accurate_2017}, leveraging multiple light sources~\cite{abdallah_adaptive_2011} or photodiodes~\cite{warren_improving_2016}, which can increase HR accuracy~\cite{park_adaptive_2023, lee_motion_2020}.
Many of today's consumer smartwatches thus embed multiple LEDs and photodiodes (e.g., Apple Watch Series~9, Fitbit Versa~4, Garmin Fenix~7).

\begin{figure}[t]
    \centering
    \includegraphics[width=\columnwidth]{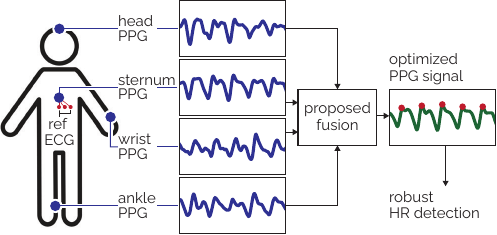}
    \caption{Our method dynamically weighs and fuses PPG signals across sensors at different body locations based on estimated signal quality, producing a continuous PPG output signal as the basis for robust HR estimation.}
    \label{fig:figure1}
\end{figure}

In this paper, we improve HR estimation through a novel multi-site PPG signal fusion method.
We build on the insight that the motion of the wearer's body affecting PPG readings is often specific to a device's location on the body, such that motion artifacts manifest to different extents in the devices worn across the body~\cite{shimazaki_effect_2018}.
Our proposed method dynamically combines PPG signals from sensors worn across the body following our window-based signal quality estimation.
For evaluation, we contribute a novel dataset of PPG recordings of 14\,participants outside the lab and over a whole day, where each participant wore four small devices with PPG sensors and one ECG sensor for reference during real-world activities (Fig.\,\ref{fig:figure1}).
Our signal fusion method robustly estimates each participant's HR, achieving a 46\% lower error on the ground-truth ECG-based HR levels than the best PPG-based HR, which originated from the head-worn sensor.

\section{Background}

Sensor fusion has frequently been used for wearable physiological sensing~\cite{mendes_jr_sensor_2016}.
Raiano et al. removed motion artifacts from breathing signals using multiple piezoresistive textile sensors, fusing signals via independent component analysis (ICA)~\cite{raiano_clean-breathing_2020}.
Combining PPG and ballistocardiograms (BCG) has also been shown to better estimate R-R intervals~\cite{antink_multimodal_2014} and HR~\cite{carek2018naptics}.
Bieri et al. jointly analyzed accelerometer and PPG signals for robust and uncertainty-aware heart rate estimation~\cite{uai2023-beliefppg}.
For HR estimation, past work has also investigated sensor fusion.
Wartzek et~al. separately detected heartbeats in PPG and capacitive ECG signals acquired in multiple locations on a mattress and then combine them for better estimation~\cite{wartzek_robust_2014}.
Lee et~al. combined PPG signals at multiple wavelengths using ICA to reduce motion artifacts~\cite{lee_motion_2020}, whereas Holz et al. combined multiple PPG sensors across the head to derive HR~\cite{holz_glabella_2017}.
Similarly, our proposed method fuses multiple PPG signals into a single output signal, albeit sourced from locations across the wearer's entire body to robustly estimate HR.

\section{Methods}

\begin{figure}
    \centering
    \includegraphics[width=\columnwidth]{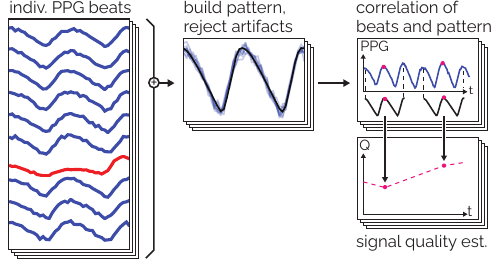}
    \caption{Our method estimates the quality of signal segments by deriving a normalized aggregate and correlating it with a leaning triangle wave to reject bad input signals during template formation.
	    This process is separate for each sensor location to account for local differences in signal morphology.}
    \label{fig:template}
\end{figure}

We propose a novel method that dynamically fuses PPG signals from multiple sensors distributed across a wearer's body based on their temporal and intermittent signal characteristics.
Our method first estimates these based on assessing individual signal quality and then weighs segments to derive the combined output signal.
Fig.\,\ref{fig:template} illustrates our method.

\subsection{PPG Peak Detection and HR Measurement}
\label{subsec:peak_and_hr}

For each PPG signal input, our method first detects segments by identifying systolic peaks in each data stream.
We first apply a bandpass filter to the input (Butterworth, 0.6-3.3\,Hz passband) and then detect peaks whenever a PPG signal crosses its moving average plus an offset, building on van Gent et~al.~\cite{van_gent_heartpy_2019}.
We determine the offset by minimizing the variance in the resulting peak intervals over a one-minute window.
Peaks are removed if the intervals between neighbors would result in a HR greater than 185\,bpm.

To reduce the impact of spurious peaks on HR computation, we only process interbeat intervals (IBI) that lie within at least five consecutive IBI where $\nicefrac{\min IBI}{\max IBI} > 0.51$.
This filtering step is relatively lenient and retains most PPG beats, even less accurate sections that more conservative filters would remove.
Our method explicitly preserves these segments to ensure the presence of signals over the full time of captured data across sensors and because they are input into our fusion.

Finally, HR values are estimated from the IBI over a 30-second moving window over the data.

\subsection{Signal Quality Estimation}
\label{subsec:signal_quality_estimation}

To assess a PPG segment's signal quality on a per-beat basis, our method matches them against a template, following the approach by Warren et~al.~\cite{warren_improving_2016} as shown in Fig.\,\ref{fig:template}.
We perform this separately for each signal to account for differences in signal morphology depending on the sensor's location on the body. 
For template formation, at each detected peak in the PPG signal, a section is used that is delimited by its neighboring peaks.
The template therefore spans two R-R intervals, increasing the likelihood of rejecting corrupt signal segments.
Sections with lengths corresponding to a HR $<40$ or $>185$ are ignored.
Each resulting section is down-sampled to 40 samples, which compensates for changes in HR, and finally z-scored (i.e., standardized to zero mean and unit standard deviation).

We then derive the template by averaging those sections that sufficiently correlate with a leaning triangle wave to exclude corrupted segments ($r>.8$).
To further lower the impact of noisy waves, we iteratively remove those segments with the lowest correlation to the template until 500 remain.

Finally, we estimate the quality of a PPG signal segment at a given systolic peak by correlating it with the template, delimiting a segment by its neighboring peaks.

\subsection{Derive Robust Output Signal from Fused PPG Signals}
\label{subsec:fusion}

Before integrating all signals using their individual segments, we align cross-sensor PPG segments to account for delays in blood pulse propagation through the wearer's arterial tree.
Within a window of $\pm 150$\,ms, we temporally offset signals to align the previously detected systolic peaks across two segments.
This window is thereby small enough to prevent misaligning heartbeats even at a HR of 180.

Next, we estimate the signal quality of PPG signals for 30-second windows of the input as the mean correlation of all contained segments with the template as outlined above.
This quality estimate is then interpolated between windows to avoid abrupt changes in quality estimates, which could lead to unwanted artifacts in the fused signal.

We obtain the fused signal $s[t]$ from $n$ PPG input signals:

\[ s[t] = \frac{\sum_{i=0}^n ppg_i[t] \cdot \max (\delta, q_i[t])^6}{\sum_{j=0}^n \max (\delta, q_j[t])^6},\]

\noindent
where $0 < \delta \ll 1$ bounds the quality estimate $q[t]$ and ensures positive values in the interval (0, 1], as negative correlations with the template are insignificant. 
Our method uses a power of six to amplify better signals in the output and account for minute differences in correlation values despite  stark differences in signal quality.
We normalize the weights to ensure that output amplitudes remain independent of the number of PPG input signals and their quality estimates.

\begin{figure}
    \centering
    \includegraphics[width=\columnwidth]{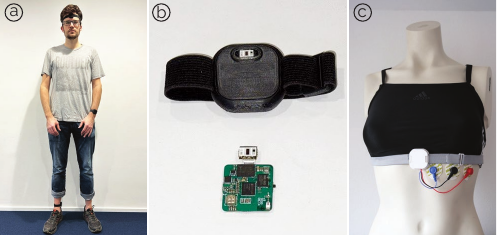}
    \caption{(a)~14 participants wore 4 standalone sensing devices for PPG at these locations for 13\,hours.
    	(b)~PPG was recorded using an SFH7072 module and a MAX86141 analog front-end.
    	(c)~The device at the sternum additionally recorded the Lead~I ECG to provide ground-truth heart rates for validation.}
    \label{fig:experiment}
\end{figure}

\subsection{Experiment Apparatus}

We evaluated our method on captured PPG signals that were recorded outside the laboratory and in real-world conditions. 
Our study collected a continuous dataset from 14 participants over the course of 13\,hours, where each participant wore four standalone devices that integrated reflective PPG sensors.
As shown in Fig.\,~\ref{fig:experiment}, devices were attached using a strap at the forehead, sternum (below clothing), ankle (supramalleolar), and wrist (dorsal).
PPG measurements were obtained using an optical analog front-end at 128\,Hz (MAX86141, Analog Devices) that connected to an optical module (SFH7072, ams-OSRAM) with a green LED (530\,nm) and a photodiode.
Sensor data was continuously read by a System-on-a-Chip (DA14695, Dialog Semi) and stored in NAND memory (TH58CYG3S0HRAIJ, Kioxia Corp.).
Each device was powered by a CR2032 coin cell battery, and all devices were synchronized by aligning recorded signals offline (with 33\,ms accuracy~\cite{meier_bmar_2023}).
For ground truth, the sternum device additionally collected the Lead~I ECG through a biopotential sensor (MAX30003, Analog Devices) that connected to gel electrodes placed on the chest.

\subsection{Experimental Protocol and Dataset}
Our dataset was recorded as follows:
Participants gathered in the morning to start the study.
An experimenter outfitted each participant with four devices and ensured PPG and ECG signal quality (20\,min).
Participants then took a minivan from Zurich to Grindelwald (140\,min), transitioned to a cablecar and train to Jungfraujoch railway station at 3460\,m above sea level (80\,min), walked through the museum and exhibition area (60\,min), walked the stairs up to the observatory (60\,min), sat down for lunch (60\,min), walked through the outside area (60\,min), rested inside (60\,min), took the train and cablecar back to Grindelwald (80\,min), and returned to Zurich on the minivan (140\,min).
Finally, the experimenter removed and collected all devices from the participants (20\,min).

Across all 14 participants, we captured $\sim$13 hours of signals from 4~PPG sensors per participant at 128\,Hz, synchronized across devices.
In addition, we captured the continuous Lead~I ECG per participant for the same duration, synchronized to the recorded PPG signals.
In total, the dataset comprises 182\,hours of synchronized signals and HR values.

\subsection{Processing}

Ground-truth HR was extracted from 30-second windows in the Lead~I reference signal, moving in steps of 5\,s and using Pan-Tompkins~\cite{pan_real-time_1985} for R peak detection.
PPG-based HR was computed as outlined above for each device (i.e., signal source) separately, and the fused PPG signal following our method was computed per person as described above.
As a baseline, ICA was applied to the PPG signals and HR was extracted from the best output following related work~\cite{lee_motion_2020}.

\section{Results}
\label{sec:results}

\setlength{\tabcolsep}{4pt}

\begin{table}[]
\centering
\begin{tabular}{lccccccc}
\multirow{2}{*}{} & \multicolumn{4}{c}{sensor sources} && \multicolumn{2}{c}{error in bpm} \\ \cline{2-5} \cline{7-8} 
configuration           & head   & sternum   & wrist  & ankle  && mean (std)           & median (std) \\
\toprule 
head                    & $\times$      &         &        &        && 4.38 (1.57) & 1.98 (1.16) \\ 
sternum                   &        & $\times$       &        &      && 5.57 (3.35) & 2.93 (2.62) \\   
wrist                   &        &         & $\times$      &        && 8.22 (3.40) & 5.75 (3.19) \\ 
ankle                   &        &         &        & $\times$      && 7.63 (4.02) & 4.62 (4.00) \\
\midrule 
ICA                     &        &         &        &        &&             &             \\ 
wearable (2)            & $\times$      &         & $\times$      &        &&  5.04 (2.22) & 1.96 (1.01) \\ 
wearable (3)            & $\times$      &         & $\times$      & $\times$      &&  5.72 (2.51) & 1.75 (0.94) \\ 
all                     & $\times$      & $\times$       & $\times$      & $\times$      && 5.01 (2.58) & 1.53 (0.77) \\
\midrule 
\multicolumn{4}{l}{\textbf{Ours}}                      &        &&             &             \\ 
wearable (2)            & $\times$      &                & $\times$      &               &&  3.52 (1.90) & 1.47 (1.30) \\ 
wearable (3)            & $\times$      &                & $\times$      & $\times$      &&  3.01 (1.44) & 0.99 (1.04) \\ 
all                     & $\times$      & $\times$       & $\times$      & $\times$      &&  \textbf{2.37 (1.14)} & \textbf{0.53 (0.48)}
\end{tabular}
\vspace{1mm}
\caption{Errors in HR calculated from PPG signal combinations.}
\label{tab:results}
\end{table}

\begin{figure}[b]
    \centering
    \includegraphics[width=\columnwidth]{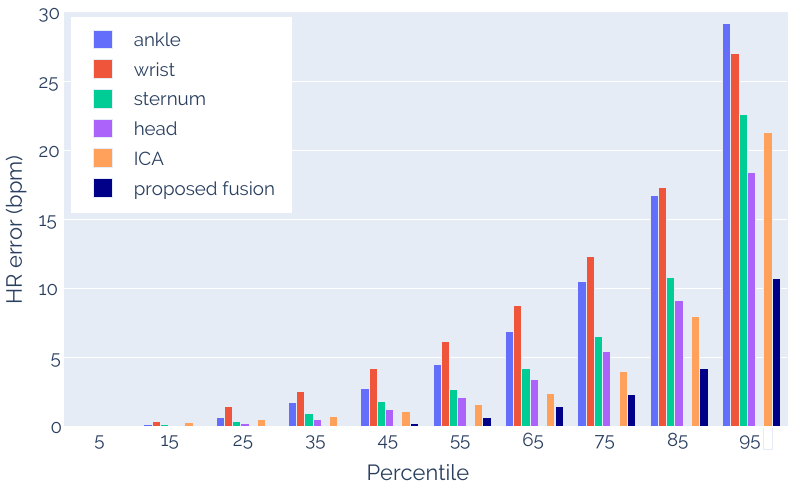}
    \vspace{-7mm}
    \caption{Percentiles of HR segments sorted by error, comparing mean error by device placements and aggregation method across all 14 participants.}
    \label{fig:percentile}
\end{figure}

Across all participants, the HR values based on the single PPG signal recorded from the head was most accurate (mean error 4.38\,bpm), followed by the sternum (5.57\,bpm), ankle (7.63\,bpm), and wrist (8.22\,bpm).
Error magnitudes thereby plausibly correspond to the amount of motion experienced by each device and thus body site during the outside activities.

We evaluated the performance of the proposed fusion method and ICA in three configurations as shown in Tab.\,~\ref{tab:results} across four, three, and two devices, respectively.

\subsection{Cross-sensor Signal Fusion}

The proposed signal fusion method overall produced HR values with the lowest mean and median error.
All three combinations resulted in a lower mean and median HR error than obtaining HR values from any of the sensor devices alone.

Compared to the HR calculated from the best-performing single device (i.e., head), the mean error decreased by 20--46\%, whereas the median HR error was 26--73\% lower.
As shown in Fig.\,\ref{fig:percentile}, this substantial reduction in HR error is not just present during phases of high HR error in individual PPG signals (higher percentiles, right);
it is also evident when the HR error in the contributing PPG signals is already low (lower percentiles, left), which highlights the potential of our method across signal quality ranges.

Fig.\,\ref{fig:percentile} additionally juxtaposes the error reduction for multi-site signal integration (fusion, ICA) and single-site PPG-based HR calculation.
This further illustrates the usefulness of multi-site PPG for robust HR calculation during motion and activity.

\begin{figure}
    \centering
    \includegraphics[width=\columnwidth]{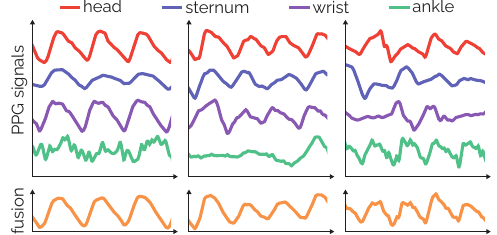}
    \caption{Three examples of four PPG signals from different body locations and the fused output signal produced by our method.}
    \label{fig:examples}
\end{figure}

\subsection{ICA}

Comparing the best signal output of ICA with the individual device placements, in no setting it produced a lower mean error in HR than using the best-performing input PPG from the head.
However, this signal from ICA-based processing did result in HR with a lower \emph{median} error than HR from any single PPG.
This suggests that it can be useful for more robust HR detection in most cases, but it also indicates that its output is more severely corrupt with higher-error HR calculation when ICA does not output a good PPG signal.

\section{Conclusion}

We have introduced a novel method for robustly estimating HR from PPG signals captured in real-world settings.
Our method takes as input PPG from multiple sensors, distributed across the wearer's body, and dynamically assesses the quality of individual samples before synthesizing a fused signal using weighted input portions.
Our method is agnostic to the number of PPG sources or body sites where sensors are worn, and its computational efficiency allows it to scale to a number of distributed sensors.
Through dynamic signal integration and synthesis, our method implicitly leverages moments of lower motion and thus signal artifacts for any sensor to represent such portions more dominantly in the fused output.

We evaluate our method on a novel dataset that we captured from 14 participants over 13 hours during outdoor and mountain activities.
Four small wearable devices distributed over the body continuously recorded PPG signals, synchronized with a Lead~I ECG captured for ground-truth HR calculation for each participant.
Our fusion method predicted HR values with a mean error that is 71\% lower than PPG-based HR from the wrist (91\% lower median error even), and 46\% lower mean error than HR from the forehead.
The fusion benefits of our method remain even for combinations of wrist+head (e.g., smartwatch+smartglasses), where the mean error is still 20\% lower than on just the head (26\% lower median error).
Fig.\,\ref{fig:examples} shows a qualitative assessment of how our method recovers PPG \emph{morphology} with examples from our dataset when individual HR estimation had a high error.
We believe that our results can inspire further research on distributed wearable health monitoring systems in the context of body sensor networks and physiological sensing to further advance reliable and continuous physiological monitoring in the wild outside controlled laboratory conditions.

\bibliographystyle{IEEEtran}
\bibliography{manual_refs}
\balance{}

\end{document}